\documentclass[aps,pra,floatfix,superscriptaddress,twocolumn,showpacs,10pt]{revtex4-1} \usepackage{amssymb, amsmath,color,mciteplus,graphicx,subfigure}
\usepackage{times}

\begin{document}

\title{Universal holonomic quantum gates in decoherence-free subspace on superconducting circuits}

\author{Zheng-Yuan Xue} \email{zyxue@scnu.edu.cn}
\affiliation{Guangdong Provincial Key Laboratory of Quantum Engineering and Quantum Materials,  and School of Physics\\ and Telecommunication Engineering, South China Normal University, Guangzhou 510006, China}

\author{Jian Zhou}
\affiliation{Guangdong Provincial Key Laboratory of Quantum Engineering and Quantum Materials,  and School of Physics\\ and Telecommunication Engineering, South China Normal University, Guangzhou 510006, China}
\affiliation{National Laboratory of Solid State Microstructure, Nanjing University, Nanjing 210093, China}

\author{Z. D. Wang} 
\affiliation{Department of Physics and Center of Theoretical and Computational Physics,  The University of Hong Kong,\\ Pokfulam Road,  Hong Kong, China}

\date{\today}

\begin{abstract}
To implement a set of universal quantum logic gates based on non-Abelian geometric phases, it is a conventional wisdom that  quantum systems beyond two levels are required, which is extremely difficult to fulfill for superconducting qubits and appears to be a main reason why only single-qubit gates were implemented in a recent experiment [A. A. Abdumalikov Jr. \emph{et al}.,  Nature  (London) {\bf 496}, 482 (2013)]. Here we propose to realize nonadiabatic holonomic quantum computation in decoherence-free subspace on circuit QED, where  one can use only the two levels in transmon qubits, a usual interaction, and a minimal resource for the decoherence-free subspace encoding. In particular, our scheme not only overcomes the difficulties encountered in previous studies, but also can still achieve considerably large effective coupling strength, such that high fidelity quantum gates can be achieved.   Therefore, the present scheme makes realizing robust holonomic quantum computation with superconducting circuits very promising.
\end{abstract}

\pacs{03.67.Lx, 42.50.Dv, 85.25.Cp}

\maketitle

\section{Introduction}

Under adiabatic cyclical evolution, a quantum system acquires a phase factor, which consists of both dynamical and geometric components. When the eigenstates of the system are nondegenerate, the geometric component is the well-known Berry phase \cite{berry}. For the degenerate case, it is a unitary operator acting on the degenerate subspace, i.e., holonomy \cite{wz}. As geometric phases are determined by the global property of the evolution path, the geometric method of quantum computation has been shown to possess some built-in noise-resilience features \cite{exp1,exp2,exp3,exp4,exp5,exp6,zhunoise,noise}. In general, the holonomies do not commute with each other, and thus can be used to construct a universal set of quantum gates \cite{h,h1,h2,h3,h4,h5,zp,oo,h6,h7,h8,h9,h10}, i.e., the  holonomic quantum computation (HQC).

On the other hand, the adiabatic method of quantum computation intrinsically leads to long gate operation time, which may be comparable with the lifetime of typical qubits \cite{xbw,zhuspeed}. This motivates research on  quantum computation based on the  nonadiabatic  geometric phases. Recently, nonadiabatic HQC has been proposed using three-level $\Lambda$ systems \cite{n1} with the experimental implementation of some elementary gates \cite{e1,e2,e3,e4}. However, the excited state is resonantly coupled when implementing the quantum gates \cite{n1}, and thus, its limited lifetime is a main challenge in practical experiments. Note that this limitation may be avoided in experiments in Refs. \cite{e3} and \cite{e4} because they use the three magnetic states of a nitrogen-vacancy center in a diamond. However, for a superconducting transmon qubit, this limitation does exist, and recent experiment has verified only single-qubit gates \cite{e2}. The energy levels of a transmon qubit \cite{transmon} are in a ladder shape, and the anharmonicity is small, which limits the coupling strength between neighboring levels to the order of 10 MHz in order to individually address the interactions \cite{e2,supertime}. Therefore, even with newly demonstrated good coherent times of multilevels in the transmon qubit \cite{supertime}, the implementation of a nontrivial two-qubit holonomic gate, which needs much more complicated cavity-induced interaction between two three-level systems \cite{n1},  is still very challenging. Alternatively, there are schemes using circuits more complicated than transmons to mimic a multilevel system \cite{exp5,h5,h7}. However, this will inevitably introduce additional noises from the environment because more circuits and control elements are needed.

Meanwhile,  many efforts have also been made to combine HQC with the decoherence-free subspace (DFS) encoding \cite{dfs1,dfs2,dfs3}. HQC in DFS \cite{law,n2,n3,n4,n42,n5} can consolidate both the noise resilience of the encoding and the operational robustness of holonomies. As for transmon qubits,  this protocol is much more difficult to implement because it requires at least two transmon qubits to encode a logical qubit, and thus, more complex interactions among qubits are needed, even in the single-qubit-gate case. Moreover, previously proposed schemes  based on HQC in DFS  usually need at least three physical qubits to encode a logical qubit; note that the use of two physical qubits is a minimum resource needed, as  in Ref. \cite{xlf} for geometric entangling gates.

In this paper, we propose to implement a nonadiabatic HQC in DFS with a typical circuit QED setup. Our scheme avoids the above-mentioned difficulties. First, only the two energy levels of the transmon qubits are  involved. Second, for the single-qubit case, our implementation relies solely on the effective resonate qubit-cavity interaction. Meanwhile, for the two-qubit case, only a conventional detuned interaction is required, where the detuning between a transmon qubit and the cavity is fixed, and thus, we have plenty freedom to avoid the limitation due to the small anharmonicity of transmons.  Third, we use two transmons to encode a logical qubit, which is the minimal resource for the DFS encoding.  Therefore, our scheme presents a promising method for HQC on superconducting circuits.

\begin{figure}[tbp]\centering
\includegraphics[width=\columnwidth]{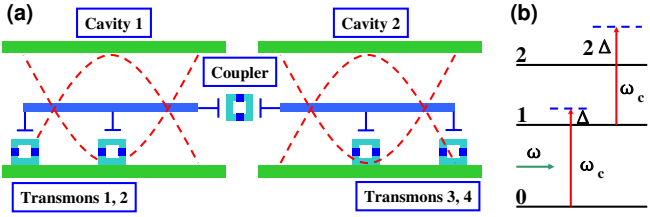}
\caption{(Color online) Illustration of the setup of our scheme. (a) The coupled two-transmission-line-resonator system for universal holonomic quantum computation in a decoherence-free subspace,  in which each resonator has two transmon qubits to encode a logical qubit. (b) The qubit and cavity coupling configuration.}
\end{figure}

\section{The setup and interaction}
The setup we consider is illustrated in Fig. 1(a), which consists of two coupled transmission line resonators (one-dimensional cavities). In each cavity, there are two transmon qubits \cite{transmon} capacitively coupled to it. This coupled system is the building block of our scheme, and the two transmon qubits encode a logical qubit. For the case of single-logical-qubit gates, we consider only the dynamics within a cavity, while the coupling between the two cavities is switched on when implementing the two-logical-qubit gate.  Furthermore, one may repeat this block to construct a one-dimensional chain of logical qubits; that is all the parameters of the odd and even qubits are the same as those of the first and second qubits, respectively.

We first detail our setup for the single-qubit case.   A transmon qubit is composed of two identical Josephson junctions in a loop configuration  and is shunted by a large capacitance. The transmon qubit is quantized  and its lowest two energy levels can be used to construct our physical qubit states with the effective Hamiltonian
$H_\text{q,j}= \omega_\text{q,j}  \sigma^{\text{z}}_j /2$
($\hbar = 1$), where $\omega_\text{q,j}$ is the energy splitting of the transmon qubit and $\sigma^{\text{z}}_j$ is the Pauli matrix of the  $j$th transmon qubit in its eigenbasis.  For typical values of experimental parameters,  $\omega_\text{q,j} \sim [4, 10]$ GHz \cite{cqed}. The transmon qubits are located at the voltage antinodes of the relevant cavity mode, $H_\text{c}=\omega_\text{c}  a^\dagger a$, where $\omega_c$, $a$, and $a^\dagger$ are the frequency, annihilation, and creation operators of the cavity, respectively. The coupled system is described by \cite{cqed}
\begin{eqnarray}  \label{jc}
H_\text{JC} =H_\text{c}+ \sum_{j=1}^n \left[H_\text{q,j}+ g(\sigma_j a^\dagger + \sigma_j^\dagger a)\right],
\end{eqnarray}
where $g$ is the qubit-cavity coupling strength (assumed to be real), $\sigma_j$ is the transmon lower operator, and $\sigma^\dagger_j=(\sigma_j)^\dagger$. Here we consider the case of $\Delta=(\omega_\text{c}-\omega_\text{q})\gg g$; that is, the qubit-photon interaction  acts perturbatively.

To get a resonate interaction between a  selected transmon qubit and the cavity, the qubit is biased by an ac magnetic flux, which will introduce periodical modulation \cite{eff} of the qubit transition frequency in the form of
\begin{eqnarray}
\omega_\text{q,j}(t)=\omega_\text{q}+ {\varepsilon_j } \sin (\omega_j t-\varphi_j).
\end{eqnarray}
This modulation may effectively turn  the qubit's sideband on resonance with the cavity frequency. This can be clearly seen by moving to the rotating frame defined by $U=U_1\times U_2$,
\begin{eqnarray}
U_1=\exp\left[-i \left( {\omega_\text{q}\over 2} \sigma_j^z
+ \omega_\text{c}  a^\dagger a \right) t \right],\notag\\
U_2=\exp\left[i \sigma_j^z  {\alpha_j \over 2 } \cos (\omega_j t-\varphi_j)\right],
\end{eqnarray}
with $\alpha_j=\varepsilon_j / \omega_j$, and the transformed Hamiltonian is
\begin{eqnarray}\label{trans}
H_{\text{trans}}&=& U^\dagger H_\text{JC, j}U-i U^\dagger \frac{\partial U}{\partial t}\notag\\
&=& U^\dagger\left (g\sigma_j a^\dagger +\text{H.c.} \right)U \notag\\
&=& U_2^\dagger\left (g\sigma_j a^\dagger e^{i \Delta t}+\text{H.c.} \right)U_2 \\
&=&  g  \sigma_j a^\dagger e^{i \Delta t}
\exp\left[i \alpha_j  \cos (\omega_j t-\varphi_j)\right]
+\text{H.c.}. \notag
\end{eqnarray}
Using the Jacobi-Anger identity of
\begin{eqnarray}
&& \exp[i\alpha_j \cos(\omega_j t-\varphi_j)]\notag\\
&=& \sum_{m=-\infty}^\infty i^m J_m(\alpha_j) \exp[im(\omega_j t-\varphi_j)], \notag
\end{eqnarray}
and $J_{-m}(\alpha_j)=(-1)^m J_m(\alpha_j)$, with $J_m(\alpha_j)$ being Bessel functions of the first kind, the transformed Hamiltonian reduces to
\begin{eqnarray}
H_{\text{d,j}} &=&g J_0(\alpha_j)(\sigma_j a^\dagger e^{i \Delta t}+\text{H.c.})\\
&+& g  \sigma_j a^\dagger \sum_{m=1}^\infty i^m J_m(\alpha_j)
e^ {i [(\Delta -m\omega_j) t + m\varphi_j]} +\text{H.c.}.\notag
\end{eqnarray}

When $\omega_j=\Delta$, the effective resonate qubit-cavity interaction will be in the form of
\begin{eqnarray}  \label{driven}
H_\text{d,j} = g_j  \sigma_j a^\dagger + \text{H.c.},
\end{eqnarray}
where $g_j=g J_1(\alpha_j)\exp(i\varphi_j+\pi/2)$, we have applied the rotating-wave approximation by neglecting the oscillating terms, and the smallest oscillating frequency is $\Delta$, i.e.,
\begin{eqnarray} \label{osc}
H_\text{osc} = g \sigma_j a^\dagger \left[J_0(\alpha_j) e^{i\Delta t}
+J_2(\alpha_j) e^{-i(\Delta t - 2\varphi_j)}\right]+ \text{H.c.}. \notag\\
\end{eqnarray}
In this way, we can have full control of the coupling strength $g_j$ by varying the externally driven ac magnetic flux, i.e., by controlling the amplitude $\varepsilon_j$ and phase $\varphi_j$. Meanwhile, assuming that the anharmonicity of the transmon is the same as $\Delta$, as the transition frequency of $|1\rangle\leftrightarrow|2\rangle$ is $\omega_q-\Delta$, the third level of the transmon can only couple dispersively with the detuning $2\Delta$, as shown in Fig. 1(b). Finally, note that the resonate interaction in Eq. (\ref{driven}), for the case with more than two qubits, has the conserved quantity of total excitation $N=\sum_{j=1}^n\sigma_j^\dagger\sigma_j +n_c$, with $n_c$ being the photon number in the cavity.

\section{Single qubit gates}

We now proceed to deal with  the holonomies for single-qubit gates in DFS. Hereafter, to avoid confusion, we refer to our physical transmon qubits as transmons and logical qubits as qubits for short. As the transmons are placed in the same cavity, they can be treated as interacting with the same cavity-induced dephasing environment. The DFS we consider here is the subspace of
\begin{eqnarray}
S_1 = \{|100\rangle, |010\rangle, |001\rangle\}
\equiv \{|0\rangle_L,|1\rangle_L,|E\rangle_L\},
\end{eqnarray}
where the subscript $L$ denotes the states belonging to the logical qubit and $|100\rangle\equiv |1\rangle_1\otimes|0\rangle_2\otimes|0\rangle_c$, i.e., they denote the states of the first and second transmons, and the cavity, respectively. Also, we use the cavity as an ancillary, and thus, only two transmons are needed to encode a logical qubit. Note that the DFS is identical to the subspace of $N=1$, which ensures that the quantum dynamics will not go out of subspace $S_1$.

In this encoding, the Hamiltonian of the quantum system  consisting of two transmons, i.e., $j\in\{1,2\}$, resonantly coupled to a cavity reduces to
\begin{eqnarray}\label{h1}
H_1&=& g_1|E\rangle_L\langle0| +g_2 |E\rangle_L\langle1| + \text{H.c.} \notag\\
&=&\xi_1\left(\sin\frac{\theta}{2}e^{i\varphi}|E\rangle_L\langle0| -\cos\frac{\theta}{2}|E\rangle_L\langle1| + \text{H.c.}\right),
\end{eqnarray}
where $\xi_1=g\sqrt{J_1(\alpha_1)^2+J_1(\alpha_2)^2}$ is the effective Rabi frequency, $\tan(\theta/2)= J_1(\alpha_1)/J_1(\alpha_2)$, and $\varphi=\varphi_{1}-\varphi_2-\pi$. In this case, we construct a $\Lambda$-type Hamiltonian in the DFS with only resonate transmon-cavity interaction, from which an arbitrary single-qubit holonomic gate can be obtained. It is worth pointing out that the two transmons do not have to possess the same frequency, as we may use two externally driven fields with different frequencies to bring them in resonance with the cavity.

In the dressed-state representation, the Hamiltonian in Eq. (\ref{h1}) can be viewed as indicating that state $|E\rangle_L$ couples with only the "bright" state $|b\rangle=\sin\frac{\theta}{2}e^{-i\varphi}|0\rangle-\cos\frac{\theta}{2}|1\rangle$, while it decouples from the "dark" state $|d\rangle=\cos\frac{\theta}{2}|0\rangle+\sin\frac{\theta}{2}e^{i\varphi}|1\rangle$. Under the action of  $H_\text{1}$, the dark and bright states  evolve according to
\begin{eqnarray}
\begin{array}{ll}
|\psi_{1}(t)\rangle_{L}=U_{1}(t)|d\rangle=|d\rangle,\\
|\psi_{2}(t)\rangle_{L}=U_{1}(t)|b\rangle=\cos(\xi_{1}t)|b\rangle
-i\sin(\xi_{1}t)|E\rangle_L.
\end{array}
\end{eqnarray}
When the condition  $\xi_{1}\tau_{1}=\pi$ is satisfied,  the dressed states undergo a cyclic evolution as
$|\psi_\text{i}(\tau_{1})\rangle\langle\psi_\text{i}(\tau_{1})|
=|\psi_\text{i}(0)\rangle\langle\psi_\text{i}(0)|$. Under this condition, the time evolution operation on the subspace \{$|d\rangle, |b\rangle, |E\rangle_{L}$\} is given by
\begin{equation}
U_{1}(\tau_{1})=\sum\limits^{2}_{i, j=1} \left[ T e^{i\int_{0}^{\tau_{1}}[A(t)-H_1]dt}\right]_\text{i,j}
|\psi_\text{i}(0)\rangle \langle\psi_\text{j}(0)|,
\end{equation}
where $T$ is the time-ordering operator and $A_\text{i,j}(t)=i\langle\psi_\text{i}(t)|\dot{\psi}_\text{j}(t)\rangle$.
In particular, when the condition $H_\text{i,j}(t)=\langle\psi_\text{i}(t)|H_1|\psi_\text{j}(t)\rangle=0$ is met, which means that there is no transition between the two time-dependent states, the evolution satisfies the parallel-transport condition. Therefore, the geometric nature of the operation is originated from the structure of the Hamiltonian instead of the slow evolution in the adiabatic case. Under these two conditions,  in the logical qubit subspace $\{|0\rangle_L, |1\rangle_L\}$, the nonadiabatic holonomic gates can be realized as
\begin{eqnarray}
U(\theta,\varphi)=
\left(
\begin{array}{cc}
 \cos\theta & \sin\theta e^{-i\varphi} \\
 \sin\theta e^{i\varphi} & -\cos\theta \\
\end{array}
\right),
\end{eqnarray}
where $\theta$ and $\varphi$ can be tuned by choosing approximate parameters of the externally driven ac magnetic fluxes. Therefore, an arbitrary single-qubit gate can be achieved.

\begin{figure}[tbp]
\includegraphics[width=\columnwidth]{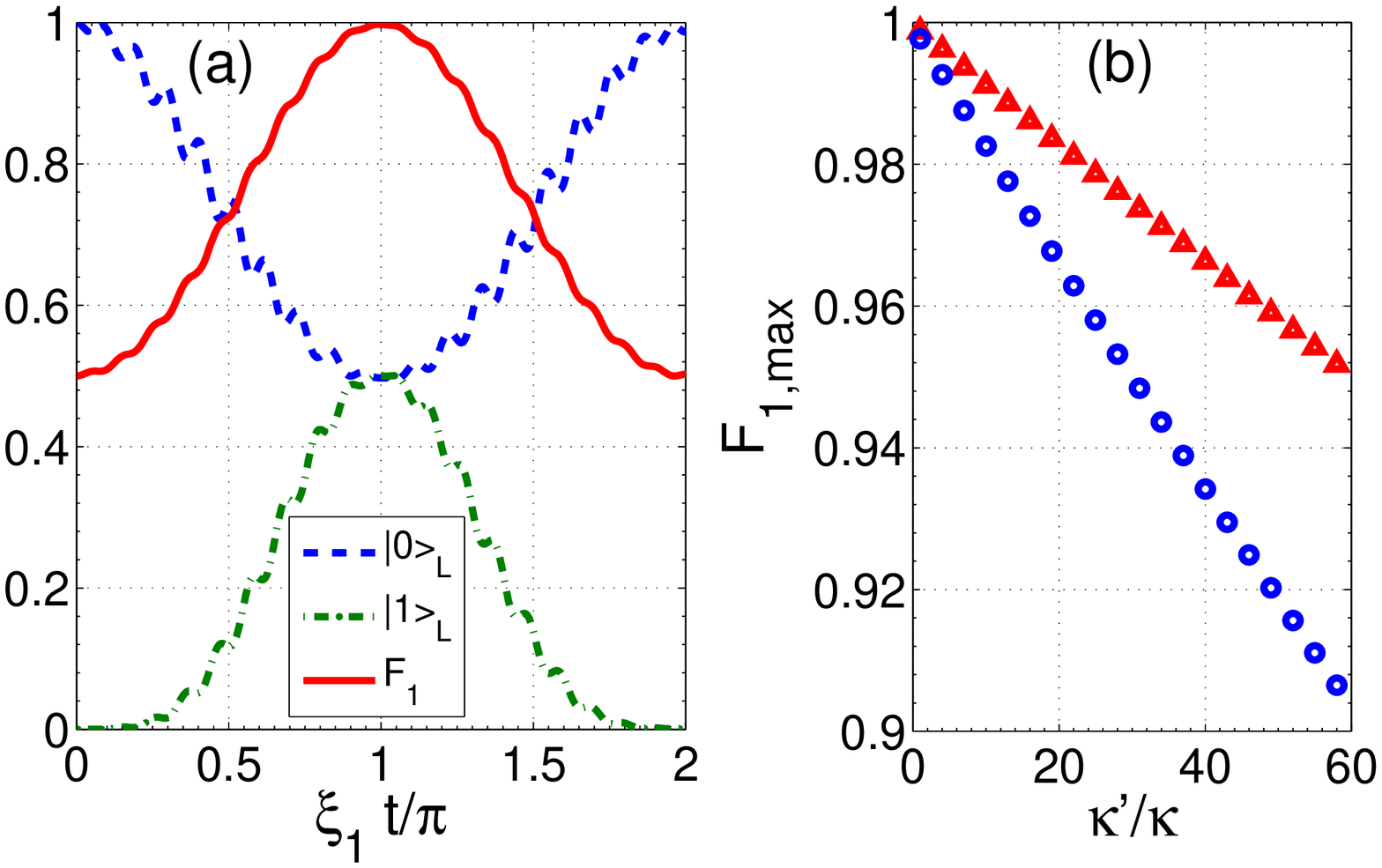}
\caption{(Color online) The performance of the proposed Hadamard gate. (a) Qubit-state population and fidelity dynamics of the Hadamard gate as a function of  dimensionless time $\xi_1 t/\pi$. (b) Maximum fidelity for different cavity decay rates $\kappa'$ (in units of  $\kappa$) and $\Delta/g=10$, with $g/(2\pi)$ being 50 and 100 MHz for the blue circles and red triangles, respectively.}
\label{fig1}
\end{figure}

The performance of  the gates can be evaluated by considering the influence of dissipation using the  quantum master equation:
\begin{eqnarray}  \label{me}
\dot\rho &=& i[\rho, H_1+H_\text{1}'] +\frac \kappa  2  \mathcal{L}(a)  \notag\\
&& + \frac{\Gamma_1}{2} \mathcal{L}(\sigma_1 + \sigma_2)
+ \frac{\Gamma_2}{2}  \mathcal{L}(\sigma_1^\text{z}+\sigma_2^\text{z}),
\end{eqnarray}
where $\rho$  is the density matrix of the considered system; $\mathcal{L}(A)=2A\rho A^\dagger-A^\dagger A \rho -\rho A^\dagger A$ is the Lindblad operator; and $\kappa$, $\Gamma_1$  and $\Gamma_2$ are the decay rate of the cavity and the decay and dephasing rates of the qubits, respectively. We have assumed that the decay and the dephasing rates of the two transmons are the same. We consider the Hadamard gate to be a typical example, where $\theta=\pi/4$ and $\varphi=0$.  To make the total coupling strong, we may choose $J_1(\alpha_1) \simeq 0.207$ and $J_1(\alpha_2) = 0.5$, which corresponds to $J_1(\alpha_1)/J_1(\alpha_2)=0.414\simeq\tan(\theta/2)$. This can be achieved by modulating  $\alpha_1=\varepsilon_1/\Delta \simeq 0.4236$ and $\alpha_2=\varepsilon_2/\Delta\simeq 1.2068$; note that tuning  the coupling strength in such a way has been experimentally demonstrated \cite{eff}. We may choose $\Delta=2\pi\times 500$ MHz, $g=2\pi\times 50$ MHz, and thus $\xi_1\simeq 0.54 g=2\pi\times 27$ MHz. To verify the approximation when obtaining Eq. (\ref{driven}), we also include some oscillating terms in the simulation. As $J_0(\alpha_1)/J_2(\alpha_1)>40$, the $J_2(\alpha_1)$ term can be safely neglected for transmon 1. Meanwhile, for transmon 2, this ratio is also larger than 4. Therefore, for demonstration purposes,  we include only the $J_0(\alpha_j)$ terms in Eq. (\ref{osc}) for the two transmons in our numerical simulation, i.e., in Eq. (\ref{me}),
\begin{eqnarray}
H_\text{1}' = \sum_{j=1}^2 \left[ g \sigma_j a^\dagger J_0(\alpha_j) e^{i\Delta t}
+ \text{H.c.}\right].
\end{eqnarray}
Quality factors of about $10^6$ have been reported for a cavity with frequencies ranging from 4 to 8 GHz \cite{cavitydecay}, and thus, the cavity decay rate $\kappa$ is on the order of kilohertz. Relaxation and coherence times of 44 and 20 $\mu$s are reported \cite{qubitdecay} for a planar transmon, which corresponds to $\Gamma_1 \simeq 2\pi \times 8$ kHz and $\Gamma_2  \simeq 2\pi \times 3.5$ kHz. As $\kappa$, $\Gamma_1$ and $\Gamma_2$ are all on the same order of magnitude, for simplicity, we treat them as if they were identical and set $\Gamma_1=\Gamma_2=\kappa=2\pi \times 10$ kHz. Suppose the qubit is initially in the state $|0\rangle_L$; we evaluate this gate by the qubit-state population and the fidelity defined by $F_1=\langle\psi_f|\rho|\psi_f\rangle$, with  $|\psi_f\rangle=(|0\rangle+|1\rangle)_L/\sqrt{2}$ being the ideally final state under the Hadamard gate. We solve numerically the master equation (\ref{me}) with $H_1'$ being given in Eq. (\ref{osc}); as shown in Fig. \ref{fig1}(a), we obtain a very high fidelity $F_1 \simeq 99.8\%$ at $t=\pi/\xi_1\simeq 18.5$ ns. Meanwhile, the higher energy levels will also be involved during the evolution; we next consider the effect of the third level of the transmons, which is the one closest to $\omega_\text{q}$. Note that more higher excited levels will not directly influence the transmon qubit subspace. For this level, we assume that the anharmonicity of the transmon is the same as $\Delta$, and thus, the cavity-induced coupling between the second and third levels when obtaining the Hamiltonian in Eq. (\ref{driven}) is dispersive, i.e., oscillating with frequency $2\Delta$. Therefore, its contribution to the infidelity of the quantum gates is similar to that of the Hamiltonian in Eq. (\ref{osc}). Assuming that the coupling strength is $\sqrt{2}g$, the infidelity that results from this third level is $0.06\%$ from our numerical simulation. In addition, when the transmons are incorporated into the cavity, its decay rate will be enhanced. Therefore, we also investigate the influence of the increase of the cavity decay on the gate fidelity; as shown by the blue circles in  Fig.  \ref{fig1}(b), for $\kappa'=30\kappa$, we can still get $F_1\simeq 95\%$. Meanwhile, the larger detuning implies $g$ can be bigger, and thus, the fidelity can be higher. As shown by the red triangles in Fig.  \ref{fig1}(b), we verify this by  choosing the detuning and transmon-cavity coupling strength to be 2$g$ and 2$\Delta$, respectively.

\section{Two-qubit gate}

At this stage, we turn to the implementation of a nontrivial two-qubit gate. To avoid the cross talk between transmons, we consider the scenario of two coupled cavities, with each one having two transmons to encode the logical qubit. In this case, a six-dimensional DFS exists
\begin{eqnarray}
S_2&=&\{|00\rangle_L=|100100\rangle, |01\rangle_L=|100010\rangle, \notag\\
&& |10\rangle_L=|010100\rangle, |11\rangle_L=|010010\rangle, \notag\\
&& |a_1\rangle=|110000\rangle, |a_2\rangle=|000110\rangle\},
\end{eqnarray}
where  $|10100\rangle\equiv|1\rangle_{1}|0\rangle_{2}|0\rangle_{c1}|1\rangle_{3}|0\rangle_{4}|0\rangle_{c2}$; $|a_1\rangle$ and $|a_2\rangle$ are two  ancillary states, both of which have two excitations within a logical qubit and thus will not be affected in the single-qubit cases (in the $N=1$ subspace).

To obtain a nontrivial two-qubit gate, we need to induce the interaction within two transmon pairs, i.e., transmons 2 and 3 and transmons 2 and 4,  but to avoid the interaction between other pairs. To achieve this, we consider that the interaction is induced by exchanging virtual photons between the two coupled cavities, the coupling of which is  \cite{cc}
\begin{eqnarray}  \label{coupled}
H_{\text{cc}} = \lambda (a_1^\dagger a_2+a_1 a_2^\dagger),
\end{eqnarray}
and the frequencies of the delocalized field modes $P_1=(a_1-a_2)/\sqrt{2}$ and $P_2=(a_1+a_2)/\sqrt{2}$ are shifted \cite{dyn} from the
bare cavity frequency as $\omega_\text{1}=\omega_\text{c}-\lambda$ and $\omega_\text{2}=\omega_\text{c}+\lambda$, respectively.
In addition, to avoid cross talk between the two interacting pairs, they have different detunings. The  setup is detailed as follows. For transmons 3 and 4, which are located in the second cavity, we do not apply externally driven on them and set $\delta=(\omega_\text{q,3}-\omega_\text{c}) =(\omega_\text{c}-\omega_\text{q,4})=2\pi\times 150$ MHz, $g_3=-g_4=g_2=g=2\pi\times 30$ MHz, and $\lambda=2\delta$. As for the third energy level,  the anharmonicity of the transmon is approximately $\Delta=2\pi\times 500$ MHz, and thus, the higher levels can be safely neglected.  Then, in the bare cavity frequency, the interaction Hamiltonian reads
\begin{eqnarray}\label{int1}
H_\text{int1}=g a_2^\dagger (\sigma_3 e^{-i\delta t} - \sigma_4 e^{i\delta t}) +\text{H.c.}.
\end{eqnarray}
Meanwhile, for transmon 2, we modulate the frequency of the  driven ac magnetic flux as $\omega_2'=2\delta$. In the rotating frame, the interaction Hamiltonian reduces to
\begin{eqnarray}\label{int2}
H_\text{int2}=g a_1^\dagger\sigma_2  \left[J_0(\beta)e^{i\delta t}
+J_1(\beta)e^{-i(\delta t-\phi)}\right]+\text{H.c.},
\end{eqnarray}
where $\beta=\varepsilon_2'/\omega_2'$ and  $\phi=\varphi_2'-\pi/2$.

However, in the presence of $H_{\text{cc}}$,
the frequencies of the delocalized cavity field modes $P_1=(a_1-a_2)/\sqrt{2}$ and $P_2=(a_1+a_2)/\sqrt{2}$ will be renormalized from the
bare cavity frequency as $\omega_\text{1}=\omega_\text{c}-\lambda$ and $\omega_\text{2}=\omega_\text{c}+\lambda$, respectively. Then, in the interaction picture, the field mode operators will be renormalized as
\begin{eqnarray}
a_1^\dagger \rightarrow {1 \over \sqrt{2}} \left( P_2^\dagger e^{i\lambda t}
+ P_1^\dagger e^{-i\lambda t}\right), \notag\\
a_2^\dagger\rightarrow {1 \over \sqrt{2} }\left( P_2^\dagger e^{i\lambda t}
- P_1^\dagger e^{-i\lambda t} \right).
\end{eqnarray}
Therefore, the total interaction Hamiltonian reads
\begin{eqnarray}
H_\text{int}&=&H_\text{int1}+H_\text{int2}\notag\\
&=&  {g \over \sqrt{2}} \left[h_1^\dagger  e^{i(\lambda+\delta) t}
+h_2^\dagger e^{i(\lambda-\delta) t}\right] +\text{H.c.},
\end{eqnarray}
where  $h_1^\dagger= P_2^\dagger [J_0(\beta)\sigma_2 - \sigma_4]
+ P_1 [J_1(\beta)e^{-i\phi} \sigma_2^\dagger -\sigma_3^\dagger]$ and
$h_2^\dagger=P_2^\dagger [J_1(\beta)e^{i\phi} \sigma_2 + \sigma_3]
+ P_1 [J_0(\beta) \sigma_2^\dagger + \sigma_4^\dagger].$ Assuming that $\{\lambda-\delta, 2\delta\} \gg g/\sqrt{2}$,  the above interaction can be treated as if it has two independent interaction channels that oscillate with distinctly different frequencies, with  the cross talk between them being suppressed by a frequency difference of $2\delta$. When $2\delta \gg g/\sqrt{2}$, the cross talk can be safely neglected. The effective Hamiltonian of the total interaction is
\begin{eqnarray}\label{twoqubit}
H_\text{eff}&=&\eta \left[J_1(\beta)e^{i\phi} \sigma_2 \sigma_3^\dagger -J_0(\beta)\sigma_2 \sigma_4^\dagger \right]  +\text{H.c.},
\end{eqnarray}
where $\eta=g^2 \lambda / (\lambda^2-\delta^2)$ and the Stark shift term has  been neglected. In order to turn off this coupling \cite{td1,td2,td3,td4}, we may modulate the coupling strength to be time dependent as $\lambda(t)=2\lambda\cos\omega t$, as recently demonstrated experimentally \cite{ccexp1,ccexp2}, and the two cavities have a  frequency    difference of $\omega=|\omega_\text{c1}-\omega_\text{c2}|$, with $\omega_\text{c1}$ and $\omega_\text{c2}$ being the resonant frequencies of the first and second cavities, respectively.

In subspace $S_2$, the Hamiltonian in Eq. (\ref{twoqubit}) becomes
\begin{eqnarray}\label{h2}
H_2&=& \xi_2\left[\sin\frac{\vartheta}{2} e^{i\phi} (|a_1\rangle_L\langle00|+|11\rangle_L\langle a_2|)\right.\nonumber\\
&&\left.-\cos\frac{\vartheta}{2}(|a_1\rangle_L\langle01|+|10\rangle_L\langle a_2|) +\text{H.c.}\right],
\end{eqnarray}
where $\tan(\vartheta/2)=J_1(\beta)/J_0(\beta)$ and the effective Rabi frequency $\xi_2=\eta \sqrt{J_0(\beta)^2+J_1(\beta)^2}$.  The effective Hamiltonian in Eq. (\ref{h2}) can be divided into two commuting parts as $H_2=H_a+H_b$, with
$$H_a= \xi_2\left[\sin\frac{\vartheta}{2}e^{i\phi}|a_1\rangle_L\langle00|
-\cos\frac{\vartheta}{2}|a_1\rangle_L\langle01| \right]+\text{H.c.},$$
$$H_b=\xi_2\left[ \sin\frac{\vartheta}{2}e^{-i\phi}|a_2\rangle_c\langle11| -\cos\frac{\vartheta}{2}|a_2\rangle_c\langle10| \right] +\text{H.c.}$$
When $\xi_2\tau_2=\pi$, the evolution operator in our logical qubit subspace reduces to
\begin{eqnarray}
U(\vartheta,\phi)=\left(
  \begin{array}{cccc}
    \cos\vartheta & \sin\vartheta e^{-i\phi} & 0 & 0 \\
    \sin\vartheta e^{i\phi} & -\cos\vartheta & 0 & 0 \\
    0 & 0 & -\cos\vartheta & \sin\vartheta e^{-i\phi} \\
    0 & 0 & \sin\vartheta e^{i\phi} & \cos\vartheta \\
  \end{array}
\right).\notag\\
\end{eqnarray}
We can see that the gate in subspace $\{|00\rangle,|01\rangle\}$ is different from the one in subspace $\{|10\rangle,|11\rangle\}$. Therefore, in general, this is a nontrivial two-qubit gate.

\begin{figure}[tbp]\centering
\includegraphics[width=0.95\columnwidth]{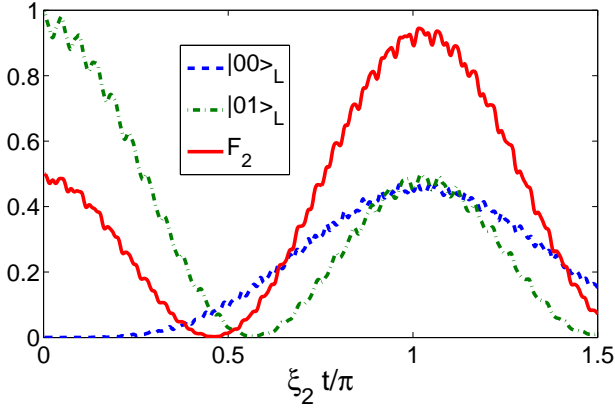}\label{f2}
\caption{(Color online) Qubit-state population and fidelity dynamics of the $U(\pi/4, 0)$ gate as a function of dimensionless time  $\xi_2 t/\pi$.}
\end{figure}

For example, when $\vartheta=\pi/4$ and $\phi=0$, it reduces to
\begin{eqnarray}
U\left({\pi \over 4}, 0\right)={1 \over \sqrt{2} } \left(
  \begin{array}{cccc}
    1 & 1 & 0 & 0 \\
    1 & -1 & 0 & 0 \\
    0 & 0 & -1 & 1 \\
    0 & 0 & 1 & 1 \\
  \end{array}
\right).
\end{eqnarray}
In this case $J_1(\beta)/J_0(\beta)=\tan(\pi/8)$, which leads to $\beta\simeq 0.77$, $J_0(\beta)=0.86$, $J_1(\beta)=0.36$, and thus $\xi_2\simeq 2\pi\times 6.2$ MHz. In addition, $J_1(0.77)/J_2(0.77)>5$ and $J_2(0.77)\gg J_n(0.77)$ for $n>2$. Moreover, the smallest oscillating frequency is $3\delta$ when $n=2$, and thus, the coupling between two transmons induced by the $J_2(0.77)$ term is much smaller in strength compared with that induced by the $J_1(0.77)$ term. Therefore, the higher-order terms are neglected in Eq. (\ref{int2}). For the initial state $|01\rangle$, we simulated the performance of this gate using the master equation with the total Hamiltonian $H_\text{t}=H_\text{cc}+H_\text{int1}+H_\text{int2}$ in Eqs. (\ref{coupled}),  (\ref{int1}),  and (\ref{int2}), as shown in Fig. 3, where a high fidelity of $F_2\simeq 94.5\%$ can be reached with the decay rates being the same as in the single-qubit case. The infidelity mainly comes from both the decoherence of the system  and the validity of the effective Hamiltonian in Eq. (\ref{twoqubit}), which are 3\% and 2.5\%, respectively. Meanwhile, considering the influence of the third level  will additional introduce about 0.5\% infidelity. In addition, when the anharmonicity is smaller, e.g., $2\pi\times 300$ MHZ, the two-qubit fidelity will be 93.1\% with $g=\delta/5=2\pi\times 20$ MHz. Finally, as transmon 3 has a different frequency than the others, its coupling should not be the same as the others. This problem can be solved by redefining $\tan(\vartheta/2)=g_3J_1(\beta)/[gJ_0(\beta)]$.

\section{Conclusion}
In conclusion,  we have proposed to implement HQC in DFS with the typical circuit QED, in which only  two levels from conventional transmon qubits are required.


\acknowledgements
This work was supported by the NFRPC (Grants No. 2013CB921804 and No. 2011CB922104), the PCSIRT  (Grant No. IRT1243), and  the fund from NJU (Grant No. M28015).

\end{document}